**Interdiffusion between gadolinia doped ceria and yttria stabilized zirconia in solid oxide fuel cells: experimental investigation and kinetic modeling**

Huixia Xu[a,b,†], Kaiming Cheng[b,c,†], Ming Chen[b,*], Lijun Zhang[a,*], Karen Brodersen[b] and Yong Du[a]

[a] *State Key Laboratory of Powder Metallurgy, Central South University, Changsha 410083, China*

[b] *Department of Energy Conversion and Storage, Technical University of Denmark, Risø campus, Roskilde 4000, Denmark*

[c] *Shandong Provincial Key Laboratory of High Strength Lightweight Metallic Materials, Advanced Materials Institute, Qilu University of Technology (Shandong Academy of Sciences), Jinan 250014, China*

[*] Corresponding author: minc@dtu.dk; lijun.zhang@csu.edu.cn

[†] Both authors contributed equally to this work.

**Abstract**

Interdiffusion between the yttria stabilized zirconia (YSZ) electrolyte and the gadolinia doped ceria (CGO) barrier layer is one of the major causes to the increment of ohmic resistance of solid oxide fuel cells (SOFCs). We present in this work experimental investigations on CGO-YSZ bi-layer electrolyte sintered at 1250 ˚C or 1315 ˚C and element transport as a function of sintering temperature and dwelling time. In order to quantitatively simulate the experimental observations, the CALPHAD-type thermodynamic assessment of the CGO-YSZ system is performed by simplifying the system to a $CeO_2$-$ZrO_2$ quasi-binary system, and the kinetic descriptions (atomic mobilities) are constructed based on critical review of literature data. The CGO-YSZ interdiffusion is then modeled with the DICTRA software and the simulation results are compared with the experimental data under different sintering or long-term operating conditions. The corresponding ohmic resistance of the bi-layer electrolyte is predicted based on the simulated concentration profile. The results implies that the interdiffusion across the CGO-YSZ interface happens mainly during sintering at high temperature, while during long-term operation at relatively lower temperature the impact of interdiffusion on cell degradation is negligible.

# 1. Introduction

Solid oxide fuel cells (SOFCs) are electrochemical energy converters, which can convert chemical energy of fuels into electrical energy directly [1–4]. Owing to high efficiency and environmental friendliness, this type of fuel cell has been investigated all over the world over the past decades. Recently, the research efforts devoted to the SOFC technology focus on cost reduction and improvement on performance, durability and reliability [5]. Several strategies have been proposed, among which reducing the operating temperature of SOFCs to the intermediate temperature range of 600 ˚C~800 ˚C attracts most of the research efforts. Under such circumstance, suitable materials for the so-called intermediate temperature SOFCs (IT-SOFC) have been investigated elaborately. Owing to its high performance and good stability, strontium and cobalt co-doped lanthanum ferrite (LSCF) is recommended as the promising candidate material for the cathode [5–7]. Unfortunately, the LSCF cathode reacts with the yttria stabilized zirconia (YSZ) electrolyte, forming insulating zirconate phases ($La_2Zr_2O_7$ or $SrZrO_3$) at the cathode-electrolyte interface [8]. To prevent undesired reactions between the LSCF cathode and the YSZ electrolyte, a reaction barrier layer made of gadolinia doped ceria (CGO) is often employed in-between to form a bi-layer electrolyte [9,10]. Several phase transformation phenomena have been identified on the cathode side of such IT-SOFCs causing lost in cell performance [11–14]: (*i*) Interdiffusion between CGO and YSZ during cell fabrication and long-term operation, resulting in a solid solution phase at the CGO-YSZ interface with increased resistivity as compared to that of CGO or YSZ; (*ii*) Sr diffusion from the LSCF cathode to the CGO-YSZ interface; and (*iii*) A secondary $SrZrO_3$ phase forms at the CGO-YSZ interface. The current work addresses only the phenomenon of the formation of CGO-YSZ solid solution by interdiffusion, whereas the others will be addressed in our future work.

The CGO-YSZ interdiffusion process results in elemental redistribution and increased resistivity. Several studies have presented the formation of solid solution phase between YSZ and CGO and the ohmic resistant depending on the mixing composition after high temperature sintering [14–21]. Tsoga et al. [16] reported that the interdiffusion phenomenon at the CGO-YSZ interface occurs already at 1200 ºC. As the temperature increases from 1300 ºC to 1550 ºC [14–19], the observed solid state reaction and interdiffusion occurring at the CGO-YSZ interface become more serious. Conductivity measurements on the mixed solid solution have been performed at various temperatures [20–22]. It was found that the ionic conductivity of the mixed solution phase decreases due to partial substitution of YSZ with CGO, reaching a minimum at the composition of $CGO_{0.5}YSZ_{0.5}$ and then increasing again to reach that of CGO for complete substitution of CGO for YSZ. The decreased conductivity of the solution phase as compared to that of pure YSZ or CGO can be explained by a decrease in the free radius through which an oxygen anion can move [23]. Due to the dramatic decrement of ionic conductivity by forming intermixed YSZ-CGO phase, even a small interdiffusion layer can significantly increase the electrolyte resistance [20]. Hence, the mixed solid solution generated by the CGO-YSZ interdiffusion is an important contributor to the increment of ohmic resistance during either cell fabrication or long-term operation.

Despite the fact that the CGO-YSZ interdiffusion is unavoidable during cell sintering, it would still be very advantageous if the interdiffusion process and its influence on the cell performance can be quantitatively modelled. Such a modelling tool will provide guidelines for optimizing cell production parameters and for cell performance prediction. Unfortunately this is not available in the literature according to the authors' knowledge. In this work, we developed a model to simulate the CGO-YSZ interdiffusion and validated the model with experimental data obtained on real SOFC cells sintered under different conditions. The CGO-YSZ interdiffusion was characterized using scanning electron microscopy (SEM) and energy dispersive X-ray spectroscopy (EDS). Based on the DICTRA (Diffusion-Controlled

TRAnsformation) software developed in the framework of the CALPHAD (CALculation of PHAse Diagram) approach [24], the interdiffusion process between CGO and YSZ was quantitatively modeled and validated against experimental data from literature and from the current work. The electrolyte resistance was then predicted from the simulated compositional profile. The current work provides a valid tool to predict the bi-layer electrolyte resistance as a function of sintering and long-term operation conditions.

## 2. Experimental

### 2.1. Sample preparation

This work focuses on the interdiffusion between the CGO barrier layer and the YSZ electrolyte. Standard halfcells produced at the Technical University of Denmark were adopted in the study directly. The halfcell consists of a ~300 μm thick NiO/YSZ support layer, a 12–16 μm thick NiO/YSZ active electrode, an 8–10 μm thick YSZ electrolyte and a 6–7 μm thick CGO ($Ce_{0.9}Gd_{0.1}O_{2-\delta}$) barrier layer. For the production of halfcells, the support layer, active fuel electrode, electrolyte and CGO barrier layer successive tape casting i.e. a multilayer tape casting (MTC) process and lamination process was applied [25]. These four MTC layers of the tapes were cut into 16 × 16 $cm^2$ pieces and co-sintered together. To explore the kinetics of the CGO-YSZ interdiffusion, two sintering temperatures (1250 °C and 1315 °C) and different sintering periods up to 200 h were employed in the current study. Five different halfcells were produced and their sintering conditions are listed in Table 1.

### 2.2. Post-mortem characterization

Cross sections of the halfcells were characterized by SEM for visualizing the microstructure and by EDS for elemental distribution and composition analysis. The samples were fractured into small pieces and were vacuum embedded in epoxy, ground, polished, and then carbon coated to eliminate surface charging during microscopy observation. The samples were characterized using a Zeiss Supra-35 SEM equipped with a field-emission gun and an energy-dispersive X-ray spectrometer (Thermo Electron Corporation). The EDS analysis was performed using the aforementioned SEM in conjunction with Noran System Six software. For backscattered electron (BSE) imaging, an accelerating voltage of 15 kV was used. For EDS analysis, an accelerating voltage of 10-15 kV was used. The NiO/YSZ support and the active electrode were excluded in the characterization, as the current work focuses the CGO-YSZ bi-layer.

## 3. Kinetic modeling

The modeling part was carried out within the CALPHAD framework and was realized by using the Thermo-Calc and DICTRA software [24]. The spirit of the CALPHAD method is to describe the thermodynamic and kinetic properties of existing phases in materials system as functions of temperature, concentration and pressure et al., and to construct self-consistent database of functional parameters. Thermo-Calc is a powerful tool for phase equilibrium, phase diagram and phase transformation calculations, whereas DICTRA is for simulation of diffusion-controlled phase transformations in multicomponent systems. In this work, the thermodynamics and phase relations of the relevant systems were established using Thermo-Calc together with the thermodynamic descriptions developed by Du et al. [26]. For kinetic modeling, a 1-D diffusion couple model was set up for simulation (shown in Fig. 1), where the YSZ electrolyte is in contact with the CGO barrier layer. The interdiffusion behavior under various conditions across the CGO-YSZ interface was then simulated using DICTRA in conjunction with both thermodynamic and atomic mobility descriptions. In the present work, the mobility functions were developed by critical assessment of the experimental diffusivity data reported in literature using the DICTRA software. The interrelations between atomic mobility and diffusivity implemented in the DICTRA software can be described as follows:

According to the absolute-reaction rate theory, the atomic mobility of element $k$, $M_k$, may be described by a frequency factor $M_k^0$ and an activation enthalpy $Q_k$ as [27],

$$M_k = \exp(\frac{RT \ln M_k^0}{RT}) \exp(\frac{-Q_k}{RT}) \frac{1}{RT} {}^{mg}\Omega \qquad (1)$$

where $R$ is the gas constant and $T$ is the absolute temperature. ${}^{mg}\Omega$ is a factor taking into account a ferromagnetic contribution to the diffusion coefficient. In the spirit of CALPHAD approach, the composition dependency of item $\Phi_k$ ($\Phi_k$ represents $RT \ln M_k^0 - Q_k$) can be represented with the Redlich-Kister polynomial [28]:

$$\Phi_k = \sum_i x_i \Phi_k^i + \sum_i \sum_{j>i} x_i x_j [\sum_r {}^r\Phi_k^{i,j}(x_i - x_j)] \\ + \sum_i \sum_{j>i} \sum_{k>j} x_i x_j x_k [\sum_s v_{ijk}^s \, {}^s\Phi_k^{i,j,k}] \quad (s = i, j \text{ or } k) \qquad (2)$$

where $x_i$ is the mole fraction of species $i$, $\Phi_B^i$ is the value of $\Phi_B$ for pure $i$ and thus represents the value of one endpoint in the composition space. ${}^r\Phi_B^{i,j}$ and ${}^s\Phi_B^{i,j,k}$ are binary and ternary interaction parameters, respectively. Each individual $\Phi$ parameter, i.e., $\Phi_B^i$, ${}^r\Phi_B^{i,j}$ or ${}^s\Phi_B^{i,j,k}$, may be expressed as a polynomial in temperature and pressure if necessary. The parameter $v_{i,j,k}^s$ is given by

$$v_{ijk}^s = x_s + (1 - x_i - x_j - x_k)/3 \qquad (3)$$

where $x_i$, $x_j$, $x_k$ and $x_s$ are mole fractions of elements $i$, $j$, $k$ and $s$, respectively.

The interdiffusion coefficients with $n$ as the dependent species are correlated to the atomic mobilities by [27]

$$\tilde{D}_{kj}^n = \sum_i (\delta_{ik} - x_k) \cdot x_i \cdot M_i \cdot (\frac{\partial \mu_i}{\partial x_j} - \frac{\partial \mu_i}{\partial x_n}) \qquad (4)$$

where $\delta_{ik}$ is the Kronecker delta ($\delta_{ik}=1$, if $i = k$, otherwise $\delta_{ik}=0$) and $\mu_i$ is the chemical potential of element *i*. Assuming the monovacancy atomic exchange mechanism, the tracer diffusivity $D_i^*$ relates to the atomic mobility via the Einstein relation:

$$D_i^* = RTM_i \qquad (5)$$

The above deals with bulk diffusion, whereas in polycrystalline materials the grain boundary diffusion plays also an important role. Nowadays, the contribution of grain boundary diffusion has been considered in DICTRA [29–31]. There, the grain boundary diffusion is correlated to the bulk diffusion by using the same frequency factor, but a modified bulk activation energy, as specified by the equation below [32]:

$$M^{gb} = M_0^{bulk} \cdot \exp(F_{redGB} \cdot Q^{bulk} / R / T) \qquad (6)$$

where $M^{gb}$ is the mobility in the grain boundary, $M_0^{bulk}$ and $Q^{bulk}$ are the frequency factor and activation energy in the bulk, respectively, and $F_{redGB}$ is the bulk diffusion activation energy multiplier.

The total mobility including both bulk and grain boundary diffusion is then formulated as:

$$M^{Total} = \delta/d \cdot M^{gb} + (1-\delta/d) \cdot M^{bulk} \qquad (7)$$

where $\delta$, $d$, and $M^{bulk}$ are the grain boundary thickness, the grain size as a function of time and temperature, and the mobility in the bulk, respectively. $F_{redGB}$, $\delta$, and $d$ are the input parameters for the grain boundary model in DICTRA.

## 4. Results and discussion

### 4.1. SEM-EDS results

Figure 2 presents the backscattered SEM images on the polished cross-section of all the five halfcells prepared in this work. In these images the four layers constituting the halfcell can be easily identified: from top to bottom, the CGO barrier layer, the YSZ electrolyte, the NiO/YSZ anode and part of the NiO/YSZ anode support. The contrast in the micrograph stems from the difference in chemical composition and crystallographic orientation. For the halfcell sintered at 1250 ˚C for 12 h (Fig. 2a), a gap (black line) is clearly visible at the CGO-YSZ interface, indicating poor adherence between the two layers resulting from the low sintering temperature. Increasing the sintering temperature to 1315 ˚C ensures good sintering and good adherence between the CGO barrier layer and the YSZ electrolyte, but on the other hand also promotes the interdiffusion. As shown in Figs. 2b-e, after sintering at 1315 ˚C, the original CGO-YSZ interface is replaced by an interdiffusion zone with small contrast difference to that of the neighboring CGO and YSZ layers. The sintering of the YSZ layer seems to proceed much better than that of the CGO layer. A fully dense YSZ layer is achieved already after 12 h sintering at 1250 ˚C, while a large amount of pores (open or closed) still exist in the CGO layer after the same sintering procedure. The sintering of the CGO layer is improved with increasing temperature to 1315 ˚C and prolonging the dwelling period, as reflected by the number and diameter of the pores in the CGO layer. During sintering, these pores experience a coarsening process and also tend to move towards the surface of the CGO layer. These pores provide a fast diffusion path promoting the CGO-YSZ interdiffusion as reported previously [14].

To quantify the composition profiles across the CGO-YSZ interface, EDS elemental mapping was conducted on the cross-sections of the five halfcells sintered at different conditions, focusing the CGO and YSZ layers only. Fig. 3a presents a high magnification SEM image of the CGO and YSZ layers from the halfcell sintered at 1315 ˚C for 12 hours. The interdiffusion zone in the vicinity of the CGO-YSZ interface shows a small contrast difference as compared to the rest of the YSZ and CGO layers. This contrast difference probably originates from the difference in chemical composition due to interdiffusion. Based on the SEM image, the grain size in the interdiffusion zone can be roughly estimated as in the range of 1-2 µm. The main part of the YSZ layer has bigger grains, 2-4 µm roughly, whereas for the CGO layer, the grains are not easily visible. A precise determination of the average grain size requires detailed electron backscatter diffraction (EBSD) characterization work and image analysis, which will be pursued in our future work. EDS elemental mapping was conducted over the entire area marked by the yellow square in Fig. 3a. The element concentration info was then integrated along the red line, i.e. as a line scan across the CGO-YSZ interface with the line thickness same as the width of Fig. 3a. Figure 3b plots the atomic percentages of Ce, Zr, Gd and Y along the red line from the YSZ region into the CGO layer. As expected, significant interdiffusion took place after sintering at 1315 ˚C for 12 hours. The initial compositions of CGO and YSZ can be read from the right and left end of Fig. 3b respectively, as 90 at.% Ce and 10 at.% Gd for CGO and 82 at.% Zr and 18 at. % Y for YSZ, quite close to the nominal composition of CGO10 ($Ce_{0.9}Gd_{0.1}O_{2-\delta}$) and 8YSZ ($Y_{0.16}Zr_{0.84}O_{2-\delta}$). An interdiffusion zone of ~4 µm in width with the composition changing from YSZ to CGO is clearly visible. With the microscope setup adopted in the current work, the interaction volume of the electron beam is estimated to be 1 µm in diameter. Hence, the composition change detected by EDS is a clear indication of the interdiffusion. EDS measurements were conducted also on the other four samples and the results will be presented in Section 4.3, in comparison with the simulation results.

**4.2. Thermodynamics of the CGO-YSZ system**

DICTRA modeling of the CGO-YSZ interdiffusion requires input information both on the thermodynamics and on the kinetic mobility for the here-studied Ce-Gd-Y-Zr-O system. Some information exists for the sub-systems, but not for the complete Ce-Gd-Y-Zr-O system. A simplification has therefore to be made in order to conduct the planned DICTRA simulation.

Gd and Y are the minor components in CGO and YSZ, respectively. As shown in Figure 3a, their composition profiles follow those of Ce or Zr reasonably well. In the current work, we simplified the Ce-Gd-Y-Zr-O system to the Ce-Zr-O system and the influence of Gd and Y was neglected. Another simplification is to neglect the influence of oxygen partial pressure ($PO_2$). Under the typical SOFC fabrication and operation conditions (air, 600 – 1400 ºC), $PO_2$ across the CGO-YSZ interface is rather close to that of air. The Ce-Zr-O system can be treated as the $CeO_2$-$ZrO_2$ pseudo-binary system.

The thermodynamic description of the $CeO_2$-$ZrO_2$ system used in this work was adopted from Du et al. [26]. Fig. 4a presents a calculated phase diagram of $CeO_2$-$ZrO_2$. Pure $ZrO_2$ has three polymorphs, from low to high temperature, monoclinic, tetragonal, and cubic, while $CeO_2$ has one solid phase, having the same crystal structure (fluorite) as cubic $ZrO_2$. Both YSZ and CGO have a cubic fluorite crystal structure. To mimic the thermodynamics of the CGO-YSZ system, a metastable phase diagram of $CeO_2$-$ZrO_2$ (Fig. 4b) is calculated where only the cubic phase is included. A miscibility gap exists at temperatures below ~1650 K, where the cubic phase tends to separate into two phases, one ceria rich (Cubic 2) and the other zirconia rich (Cubic 1). The solubility of $CeO_2$ in cubic $ZrO_2$ and that of $ZrO_2$ in cubic $CeO_2$ at 1315 ˚C (1588 K) are indicated in Fig. 4b using red open circles, as 33 mol. % $CeO_2$ and 33 mol.% $ZrO_2$, respectively. The interdiffusion between $CeO_2$ and $ZrO_2$ follows the diffusion route determined by the phase diagram, where Ce and Zr transport along the dashed horizontal line drawn at 1315 ºC. The diffusion directions of Ce and Zr are also indicated by the two red arrows. The parameters for the thermodynamic description of the cubic phase used in the present work are listed in Table 2.

### 4.3. Modeling of interdiffusion across the CGO-YSZ interface

As shown in Fig. 1, a CGO-YSZ diffusion couple was set up in the current work to mimic the experiments. Due to lack of experimental data, the CGO-YSZ system is further simplified to the $CeO_2$-$ZrO_2$ system in DICTRA modeling. According to the actual cell fabrication parameters [25], the thickness of the $CeO_2$ and $ZrO_2$ layers was set as 6 and 8 μm, respectively. The interdiffusion profile is determined by both thermodynamics of the $CeO_2$-$ZrO_2$ system and atomic mobilities of Ce and Zr. In this work, a constant $PO_2$ along the entire $CeO_2$-$ZrO_2$ diffusion couple is assumed. Oxygen diffusion is therefore neglected. This shall represent reasonably well the conditions in the CGO-YSZ bi-layer electrolyte during cell fabrication and long-term testing.

### 4.3.1 Atomic mobility

The atomic mobility functions of Ce and Zr in the cubic phase of the $CeO_2$-$ZrO_2$ pseudo-binary system were developed via evaluating experimental data using DICTRA. Kilo et al. [33,34] measured $^{96}Zr$ tracer diffusion in $(Zr_{1-x}Y_x)O_{2-x/2}$ single crystals with $0.15 \leq x \leq 0.48$. They also presented a critical review on the published tracer diffusion coefficients of Zr in YSZ from literature [33–38], where the linear dependence of the zirconium tracer diffusion on the yttria content has been detected within a fairly wide range of stabilizer concentration (8-32 mol% $Y_2O_3$). In the present work, we use the results of Kilo et al. [33,34] with a stabilizer concentration of 10.2 mol% $Y_2O_3$ to assess the mobility parameter for Zr, considering the similar materials selection during our sample fabrication [25]. Bekale et al. determined impurity diffusion of Ce and Gd in single- and poly-crystalline YSZ [38]. They reported that the bulk and grain boundary diffusion coefficients of Gd in YSZ are rather close to those of Ce. So far no investigation has been done on the diffusion of Zr and Ce in CGO. Considering that both $CeO_2$ and $ZrO_2$ have the same crystal structure of cubic fluorite and they are totally miscible at high temperature, we hence assume that the mobility functions of Zr and Ce in CGO are the same as those of Zr and Ce in YSZ, respectively. We further neglect the influence of Y and Gd, i.e. treat CGO and YSZ as cubic $CeO_2$ and cubic $ZrO_2$, respectively. In the current work, the mobility function of Zr in c-$ZrO_2$ was evaluated based on experimental data from Kilo et al. [33,34] obtained on 10.2 mol.% yttria stabilized zirconia. The mobility function of Ce in c-ZrO$_2$ was evaluated based on experimental data from Bekale et al. [38] obtained on 10 mol.% yttria stabilized zirconia. These mobility functions were then adopted for the c-CeO$_2$ phase as well. Fig. 5 presents the calculated tracer diffusion coefficients (bulk) of Zr and Ce in c-ZrO$_2$ (YSZ) as a function of inversed temperature, compared with the experimental data from literature [33–38]. Our calculated bulk diffusion coefficients reproduce the experimental results reasonably well. The mobility functions of Ce and Zr in the cubic phase obtained from the current work are listed in Table 2. For the terminology of these functions, the readers are referred to the paper by Andersson and Ågren [27].

### 4.3.2 Interdiffusion simulation

The grain boundary diffusion is a non-negligible contributor to the overall interdiffusion of the CGO-YSZ system, and shall be taken into account after the bulk diffusivity has been evaluated. Currently, it is not possible to introduce another mobility function to account for the grain boundary diffusion separately. As introduced before, DICTRA includes a grain boundary diffusion model which reproduces experimentally measured diffusion in polycrystalline materials well [39]. An example can be found in our previous work on modeling of Ni diffusion in ferritic stainless steels [29]. In this work, for simplicity and due to lack of experimental data, we assume a constant grain size of 1 μm, both for c-CeO$_2$ (Cubic 2) and c-ZrO$_2$ (Cubic 1). We further consider no grain growth along with the CGO-YSZ interdiffusion, and the remaining parameters in Eqs. 6 and 7 are adjusted when a good fit is obtained between our simulation and the experimental data from the current work and from literature [14,17]. The grain boundary thickness $\delta$ and the bulk diffusion activation multiplier $F$ are set as constant of $5\times10^{-10}$ m and 0.64, respectively.

Figure 6 presents the simulated composition profiles (atomic ratio of Ce/(Ce+Zr)) in comparison with the EDS linescan results from the present work (Figs. 6a-6e) or from literature (Figs. 6f and 6g) [14,17]. For the EDS results, two atomic ratios are calculated, Ce/(Ce+Zr) and (Ce+Gd)/(Ce+Gd+Zr+Y). The two profiles overlap with each other, supporting our simplification of neglecting the minor components of Gd in CGO and Y in YSZ. As shown in Fig. 6, a good agreement between the simulation and the experimental data is achieved [14,17]. A small difference can be found at the interface region. One can always find a composition jump in the simulated curve at the interface between zirconia and ceria. That is caused by component partition due to the thermodynamic description of $CeO_2$-$ZrO_2$ applied in the present work, and the composition width of the composition jump corresponds exactly to the width of miscibility gap between cubic $ZrO_2$ and cubic $CeO_2$. As shown in Fig. 4b, the miscibility gap between c-$ZrO_2$ and c-$CeO_2$ is within the composition range of 33-67 mol.% $CeO_2$ at 1315 ˚C, which is in correspondence with the composition jump of 30-70 at. % Ce in Figs. 6b-6e. Similar situation can be found for the samples sintered at 1250 ˚C for 12 h (Fig. 6a) and 1400 ˚C for 4h [14] (Fig. 6f), where the width of the miscibility gap shown in the phase diagram correspond well to the width of the composition jump at the interface. Both shortens as temperature increases. Whereas for the experimentally determined EDS results, the composition change at the interface is smoother. This is to a large extent caused by the interaction volume of the electron beam, which is approximately 1 $\mu m^3$ under the microscope setup used in the current work. We further simulated interdiffusion of a 2-step process, first at 1250 ˚C for 2 hours (cell sintering) and then at 700 ˚C for 2000 hours (cell long-term testing). The calculated composition profile is plotted in Fig. 6g together with the experimental data from [17]. According to the simulation, most of the interdiffusion flux took place within the first 2 hours of sintering at 1250 ˚C, while the long-term testing at 700 ˚C for 2000 hours made very little change to the composition profile. That is mainly because the diffusion of Ce and Zr in YSZ is very slow at 700 ˚C (about $10^{-29}$ and $10^{-26}$ $m^2$ $s^{-1}$).

### 4.3.3 Prediction of the resistance

Ohmic losses in an SOFC result from ionic conductivity through the electrolyte. Tsoga et al. [22] reported the conductivity of the $CGO_xYSZ_{1-x}$ solid solution at 700 ˚C, 800 ˚C and 900 ˚C. In this work, the data at 700 ˚C was fitted using a polynomial equation:

$$\log(\sigma) = 129.408x^6 - 343.667x^5 + 335.289x^4 - 142.568x^3 + 26.596x^2 - 4.893x - 1.629 \qquad (9)$$

where $x$ corresponds to the one in $CGO_xYSZ_{1-x}$. The original experimental data points and the fitted curve is presented in Fig. 7. The simulated composition profile from DICTRA modeling can be translated into $x$ in $CGO_xYSZ_{1-x}$. With this and the conductivity formulation (Eq. 9), the area specific resistance of the bi-layer electrolyte ($ASR_{ohm}$) can then be predicted as a function of sintering and long-term testing conditions.

$$ASR_{ohm} = \int_0^l \frac{1}{\sigma} dl \qquad (10)$$

where $l$ is the diffusion distance along the entire CGO-YSZ diffusion couple, and σ is the electrical conductivity depending on composition. Fig. 8 presents the predicted resistance for the bi-layer electrolyte after sintering at 1250 ˚C or 1315 ˚C for different duration, as well as after long-term testing at 700 ˚C. The theoretical resistance of 8 μm YSZ + 6 μm CGO (i.e. with no interdiffusion) is about 0.5 Ω cm$^2$ at 700 ºC (shown as the red diamond in Fig. 8). Sintering at 1250 ˚C for 2h doubles the electrolyte resistance, while 2 h at 1315 ˚C increases the resistance by a factor of > 3. The current results show that the interdiffusion between the CGO barrier layer and the YSZ electrolyte happens mainly during sintering and to a negligible extent during long-term operation. It is obvious that the cell sintering process has a dominant contribution to the cell ohmic resistance increase, where the long-term testing at 700 ˚C made almost no change. This is consistent with the previous experimental findings [11,14].

## 5. Conclusion

In the current work, we investigated the inter-diffusion between the CGO barrier layer and the YSZ electrolyte by means of both experimental investigation and theoretical diffusion modeling. The experimental results show a strong dependence of the interdiffusion on sintering temperature and time. We detected the major element diffusion of Ce and Gd from CGO into YSZ and Zr and Y in the opposite direction, which results in the formation of solid solution between CGO and YSZ. For quantitative modelling of the observed interdiffusion process, we developed the thermodynamic as well as atomic mobility descriptions for the $CeO_2$-$ZrO_2$ system. The interdiffusion including contribution from both bulk and grain boundary diffusion was modeled by using the DICTRA software. The simulated diffusion concentration profiles can reasonably reproduce the experimental ones. Based on the simulation results, we calculate the area specific resistance of the bi-layer electrolyte ($ASR_{ohm}$) under different sintering and long-term operation conditions. The current results show that the CGO-YSZ interdiffusion happens mainly during the sintering process, while its impact on cell degradation during long-term operation at lower temperature is negligible. The present work provides a proper account of the thermodynamics and kinetics of CGO-YSZ interdiffusion, and the results can be employed to further analysis of the degradation on the oxygen electrode side of solid oxide cells.

**Acknowledgements**

This work is supported by European Horizon 2020 - Research and Innovation Framework Programme (H2020-JTI-FCH-2015-1) under grant agreement No.735918 (INSIGHT project). Huixia Xu acknowledges the support from the China Scholarship Council (No. 201706370121) for her external stay at the Technical University of Denmark.

**Tables**

**Table 1.** Sintering conditions for the halfcells produced in the present work.

| Halfcell no. | Sintering temperature | Dwelling time |
|:---:|:---:|:---:|
| 1 | 1250 ˚C | 12 hours |
| 2 | 1315 ˚C | 12 hours |
| 3 | 1315 ˚C | 24 hours |
| 4 | 1315 ˚C | 48 hours |
| 5 | 1315 ˚C | 200 hours |

**Table 2.** Gibbs energy functions and atomic mobility fucntions for the cubic $CeO_2$-$ZrO_2$ phase in the present work. Further detail on the terminology can be found in [27].

| Gibbs energy function | | |
|---|---|---|
| $^0G^{c\text{-}CeO_2}$ | =-1116114+429.345T-72.0653TlnT-0.0040536$T^2$+583870$T^{-1}$ | (298<T<3083) |
| | =-1144303+603.429T-94.14TlnT-2.413115*$10^{33}T^{-9}$ | (3083<T<6000) |
| $^0G^{c\text{-}ZrO_2}$ | =-1125234.1+496.72262T-80TlnT | (298<T<2983) |
| | =-1151298.9+568.29043T-87.864TlnT+4.87454*$10^{33}T^{-9}$ | (2983<T<6000) |
| $^0L^{cubic}$ | =27797.69 | (298<T<6000) |
| **Atomic mobility function** | | |
| $\Phi_{Ce}^{CeO_2}$ | =-516151.8-61.37T | (298<T<6000) |
| $\Phi_{Ce}^{ZrO_2}$ | =-516151.8-61.37T | (298<T<6000) |
| $\Phi_{Zr}^{CeO_2}$ | =-434344.5-104.94T | (298<T<6000) |
| $\Phi_{Zr}^{ZrO_2}$ | =-434344.5-104.94T | (298<T<6000) |

All are in SI units. The Gibbs energy functions are taken from Du et al. [26] and the atomic mobility functions are obtained in this work.

**Figure captions**

**Fig. 1.** Schematic illustration of the bi-layer electrolyte diffusion couple consisting of the CGO barrier layer (right) and the YSZ electrolyte (left).

**Fig. 2.** Cross-sectional SEM-BSE images of SOFC half cells sintered at different conditions.

**Fig. 3.** (a) SEM image, (b) EDS mapping and (c) EDS line scan across the CGO-YSZ interface for the sample sintered at 1315 ˚C for 12 hours. The yellow box marks the region used for EDS analysis.

**Fig. 4.** (a) Phase diagram of $CeO_2$-$ZrO_2$ in air and (b) metastable phase diagram of the cubic phase in the $CeO_2$-$ZrO_2$ system.

**Fig. 5.** Calculated tracer diffusion coefficients of (a) Zr and (b) Ce in YSZ as a function of 10000/T, compared with the experimental data in the literature [29–34].

**Fig. 6.** Simulated composition profiles of the CGO-YSZ diffusion couple compared with experimental data from the present work and literature [14,17]: (a) 1250 ˚C, 12 hours; (b) 1315 ˚C, 12 hours; (c) 1315 ˚C, 24 hours; (d) 1315 ˚C, 48 hours; (e) 1315 ˚C, 200 hours; (f) 1400 ˚C, 4 hours; (g) 1250 ˚C, 2 hours + 700 ˚C, 2000 hours.

**Fig. 7.** Fitted total electrical conductivity at 700 ˚C as a function of *x* in $CGO_x YSZ_{1-x}$ for the CGO-YSZ solid solutions in comparison with the experimental data from literature [22].

**Fig. 8.** Predicted area specific resistance of the CGO-YSZ bi-layer electrolyte exposed to different sintering and long-term testing conditions.

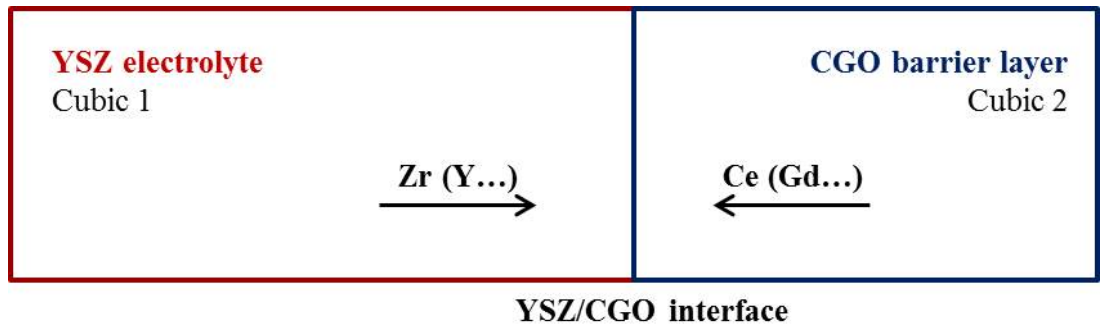

**Fig. 1.** Schematic illustration of the bi-layer electrolyte diffusion couple consisting of the CGO barrier layer (right) and the YSZ electrolyte (left).

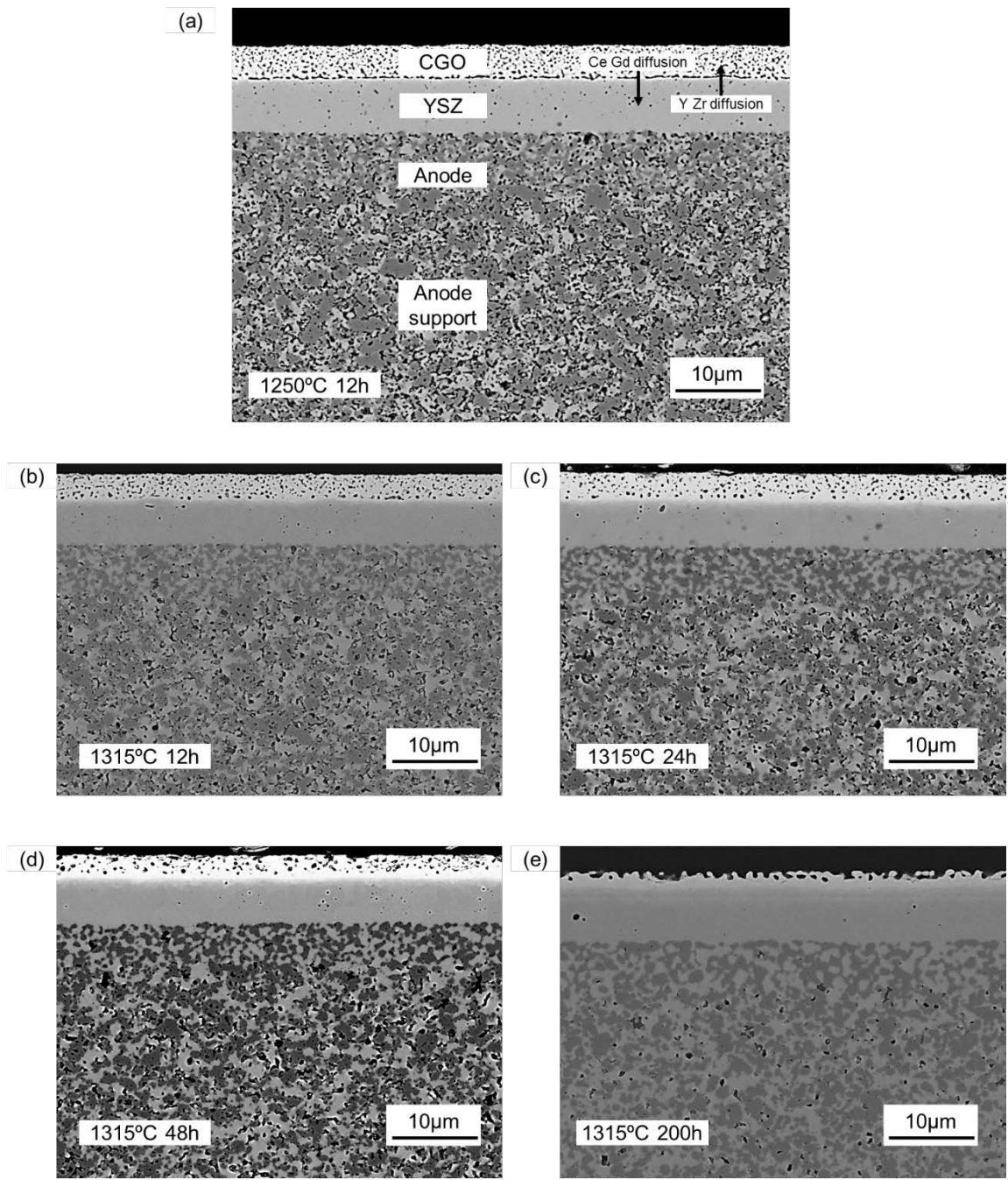

**Fig. 2.** Cross-sectional SEM-BSE images of SOFC half cells sintered at different conditions.

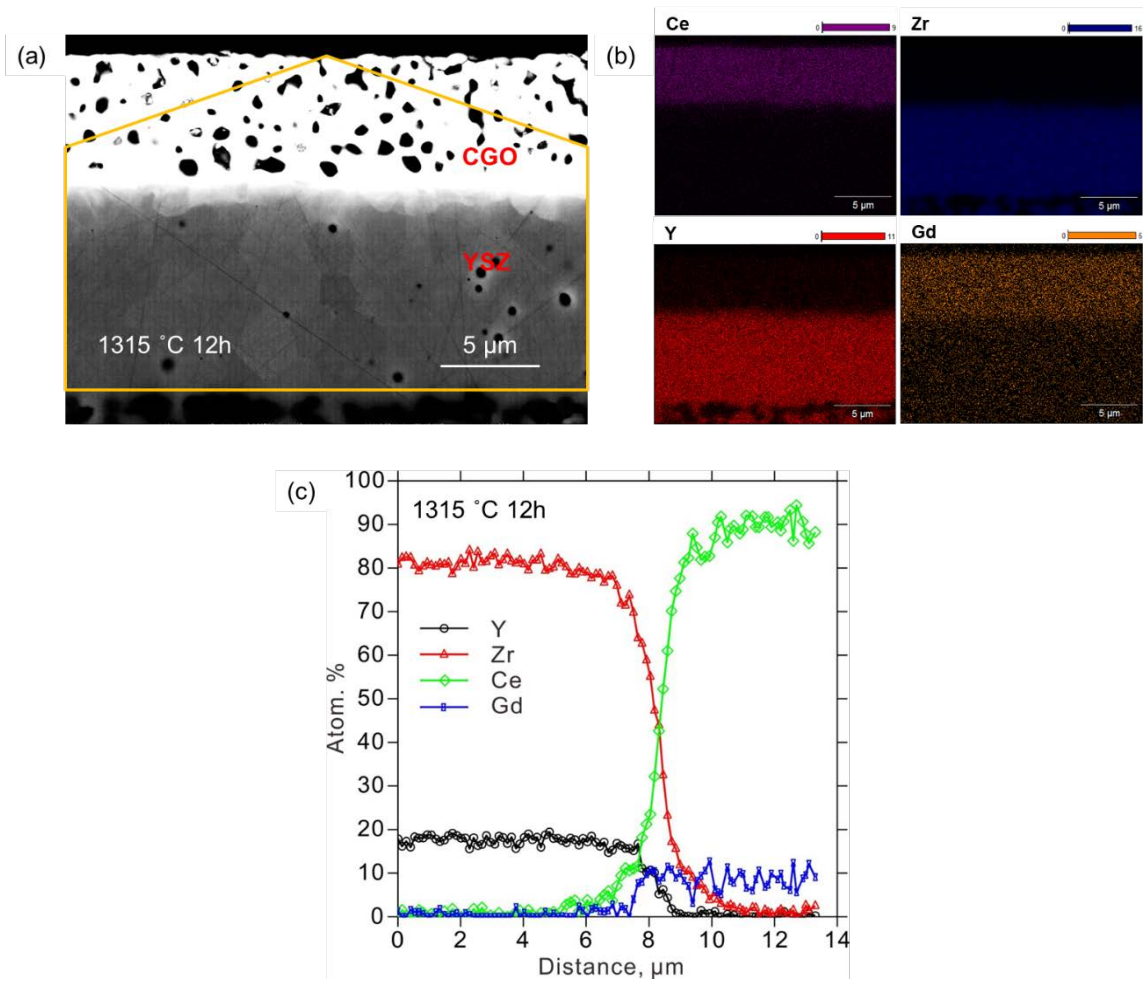

**Fig. 3.** (a) SEM image, (b) EDS mapping and (c) EDS line scan across the CGO-YSZ interface for the sample sintered at 1315 ˚C for 12 hours. The yellow box marks the region used for EDS analysis.

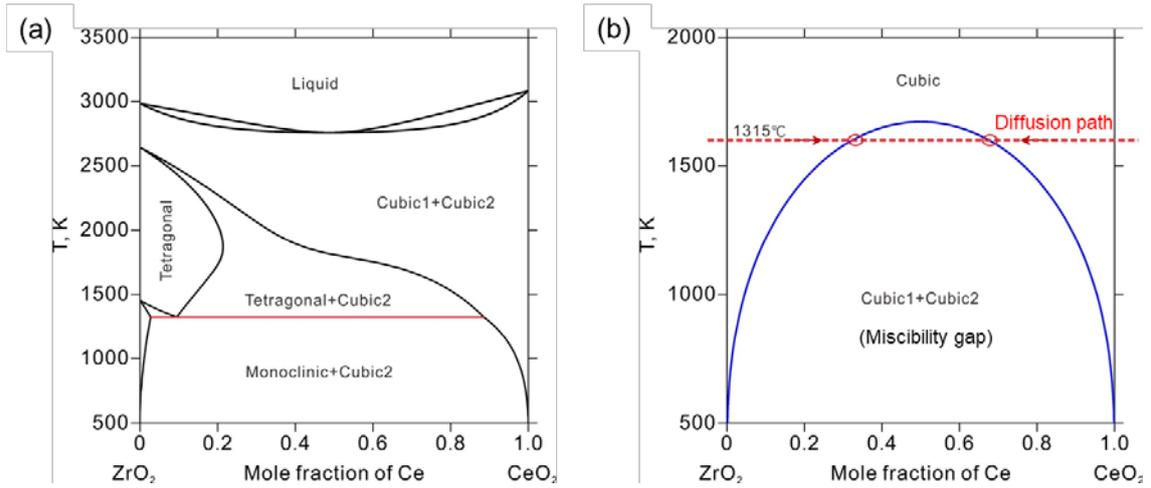

**Fig. 4.** (a) Phase diagram of $CeO_2$-$ZrO_2$ in air and (b) metastable phase diagram of the cubic phase in the $CeO_2$-$ZrO_2$ system.

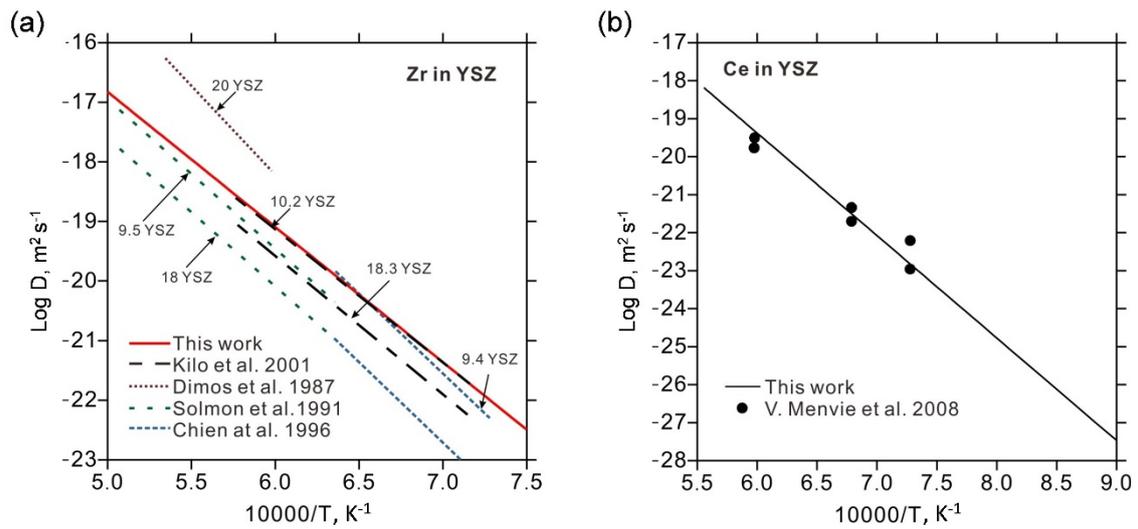

**Fig. 5.** Calculated tracer diffusion coefficients of (a) Zr and (b) Ce in YSZ as a function of 10000/T, compared with the experimental data in the literature [29–34].

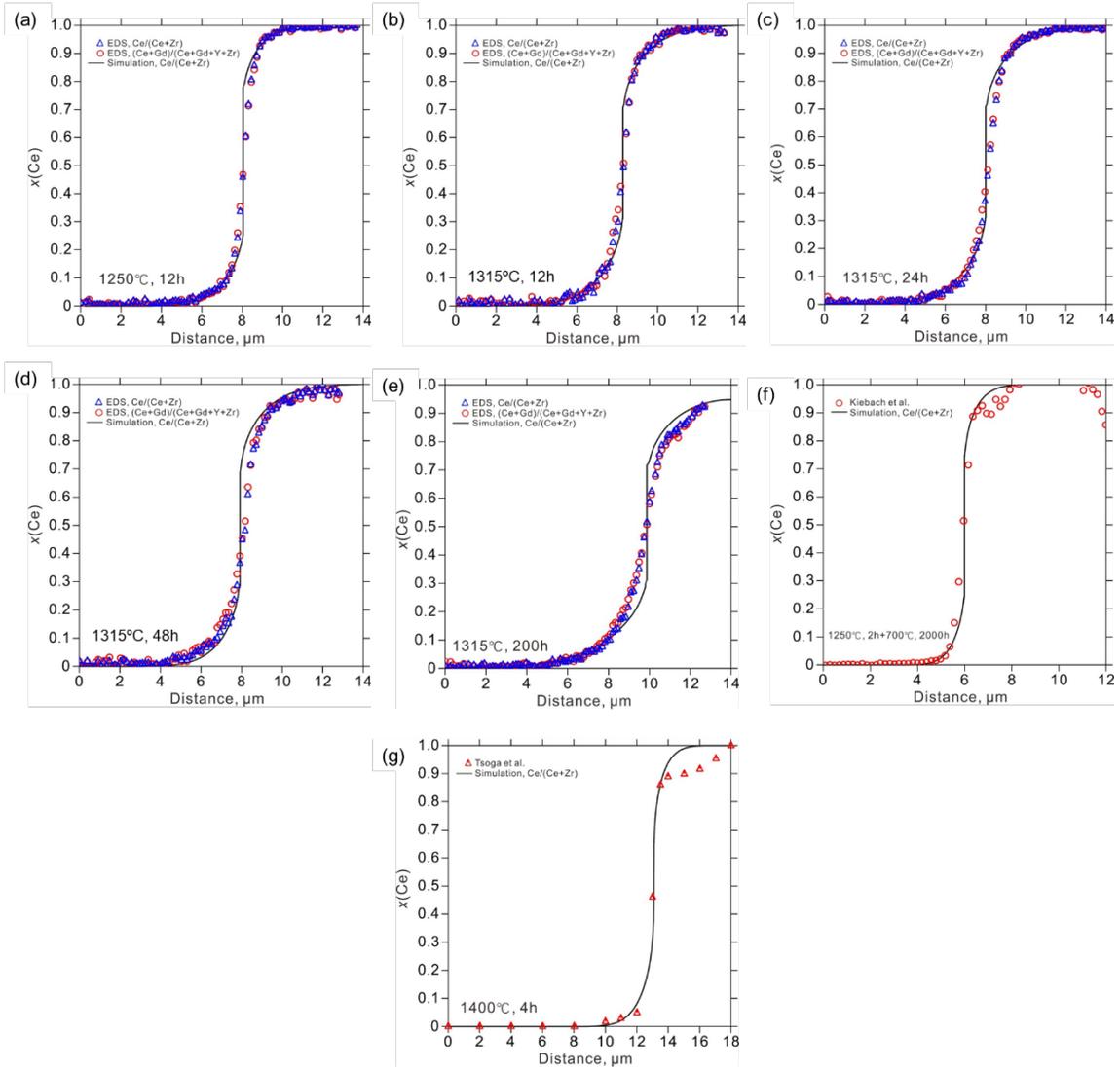

**Fig. 6.** Simulated composition profiles of the CGO-YSZ diffusion couple compared with experimental data from the present work and literature [14,17]: (a) 1250 ˚C, 12 hours; (b) 1315 ˚C, 12 hours; (c) 1315 ˚C, 24 hours; (d) 1315 ˚C, 48 hours; (e) 1315 ˚C, 200 hours; (f) 1400 ˚C, 4 hours; (g) 1250 ˚C, 2 hours + 700 ˚C, 2000 hours.

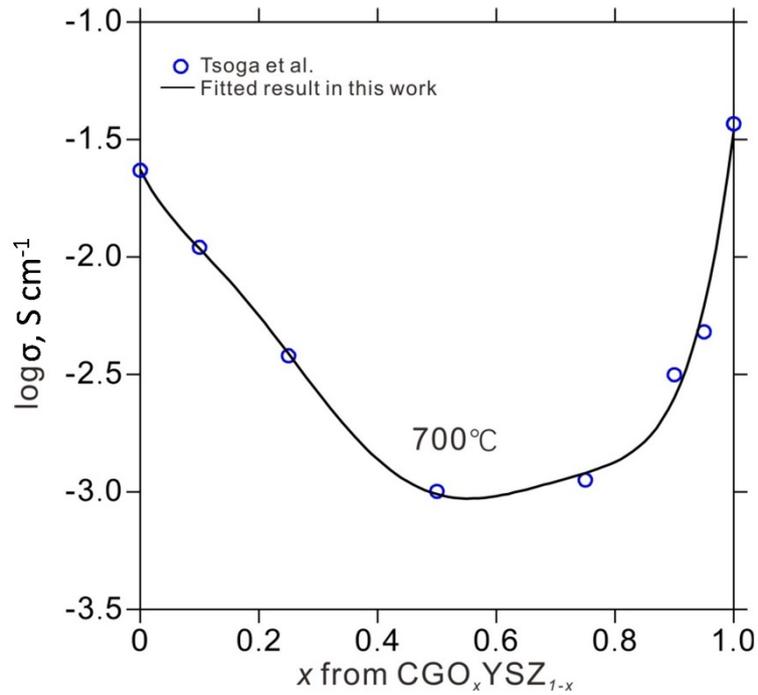

**Fig. 7.** Fitted total electrical conductivity at 700 ˚C as a function of $x$ in $CGO_xYSZ_{1-x}$ for the CGO-YSZ solid solutions in comparison with the experimental data from literature [22].

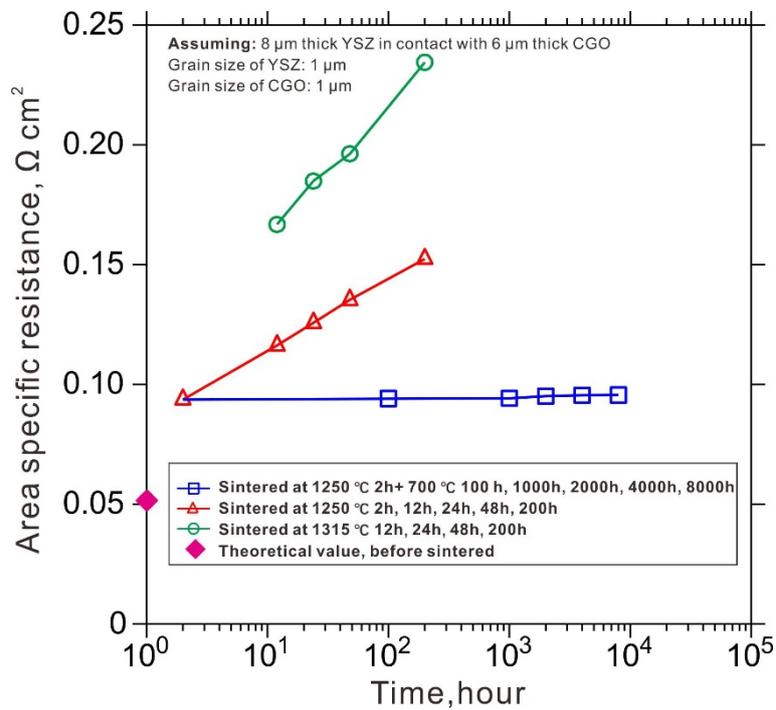

**Fig. 8.** Predicted area specific resistance of the CGO-YSZ bi-layer electrolyte exposed to different sintering and long-term testing conditions.